
\magnification=\magstep1
\hsize=5.9truein
\vsize=8.375truein
\hoffset=.45in
\parindent=25pt
\nopagenumbers
\raggedbottom
\hangafter=1
\def\makeheadline{\vbox to 0pt{\vskip-40pt
   \line{\vbox to 8.5pt{}\the\headline}\vss}\nointerlineskip}
\def\approxlt{\kern 0.35em\raise 0.45ex\hbox{$<$}\kern-0.66em\lower0.5ex
   \hbox{$\scriptstyle\sim$}\kern0.35em}
\def\approxgt{\kern 0.35em\raise 0.45ex\hbox{$>$}\kern-0.75em\lower0.5ex
   \hbox{$\scriptstyle\sim$}\kern0.35em}
\vglue1.5truein
\centerline{\bf $W_\infty$ AND ANOMALIES OF SELF-DUAL EINSTEIN THEORIES  }
\vskip22pt
\centerline{V.G.J. Rodgers}
\centerline{ Department of Physics and Astronomy}
\centerline{ The University of Iowa}
\centerline{ Iowa City, Iowa~~52242--1479}
\centerline{ April 1991 }
\vglue2.0truein
\baselineskip=12pt
\centerline{\bf ABSTRACT}
\vskip22pt
Recently it has been demonstrated that self-dual Einstein Euclidean instantons
 possess an infinite dimensional
group of symmetries which contain the standard $w_\infty$.   Since
$w_\infty$ has a central extension only in its $w_2$ subalgebra one may
claim that there is an anomaly in the symmetry group of these instantons
which is described by a three dimensional $w_2$ effective action.
Thus one may have a bosonization of the  ${3/2}$ spin anomaly of the
Eguchi-Hanson
instantons.
In analogy to $w_\infty$
we construct the three dimensional effective action for the
Lone Star $W_\infty$ algebra.  This suggests that there is a quantum
deformation of the  instantons that contributes to the effective
action of four dimensional self-dual gravity.  We comment on this possibility.
\vfill
\eject
\headline={\tenrm\hfil\folio}
\baselineskip=16pt
\pageno=1
Two dimensional conformal field theory [1] has firmly rooted itself into the
theoretical  understanding of many different physical phenomena.
For some time now there has been evidence that suggests that self-dual gauge
theories in four dimensions can be thought of as self-dual theories in two
dimensions [2].
In Ref. [3], Park is able to
show that 4 dimensional self-dual gravitational systems have the symmetries of
2D conformal theories.  In particular, four dimensional self-dual Einstein
theories which admit a single rotational Killing vector have the symmetry of
$w_\infty$.  Since the standard  $w_\infty$ of Bakas[4] has a
central extension only in its $w_2$ subalgebra it appears that the three
dimensional
$w_2$ effective action of [5] is the anomalous contribution for
Eguchi-Hanson [6] effective actions.  Furthermore different coadjoint orbits of
the $w_2$ algebra correspond to different isotropy groups of the self-dual
Einstein
equations.  The isotropy groups of the various covectors are used to construct
different solutions to the self-dual field equations.

Just as in WZW and Polyakov gravity one may think of these effective actions as
 arising from functionally integrating chiral fermions coupled to
gauge fields.  Atiyah has shown, using index
theorems, that the Dirac equation in the presence of an Eguchi-Hanson
instanton has index zero.  This implies that the chiral Dirac operator
with a self-dual spin connection, has no anomaly.
However, as shown by Hanson and R\"omer, the Rarita-Schwinger operator does
have a non-zero index,
 revealing a ${3 \over 2}$ spin anomaly in the presence of these instantons
[7].  The fact that these instantons should contribute to  the  ${3 \over 2}$
spin axial anomaly as opposed to the  ${1 \over 2}$ spin axial anomaly has been
attributed to supersymmetry [8].  All this strongly suggest that the three
dimensional $w_2$ effective action is the response of the ${3 \over 2}$ spin
fermionic measure to the underlying 2D conformal symmetry in self-dual $D=4$
theories.
These actions suggest a bosonization for the spin ${3\over 2}$ fields coupled
to self-dual connections.

We would like to
extend our  analysis of $w_2$ to the symplectic
geometry of the  ``Lone Star'' $W_\infty$ [9] case.  Throughout we will make a
distinction between the $w_\infty$, due to Bakas [4] and Bilal [10],  and
$W_\infty$, the ``Lone Star'' algebra.  The $W_\infty$ algebra corresponds to a
linear deformation of the standard $w_\infty$ algebra in such a way that all
higher
conformal spin fields have central extensions but the generators themselves do
not have definite conformal spin.

Recall that the standard $w_\infty$ algebra [4,10] can be written as
$$
\{ f,g \} =\partial_xf\partial_yg-\partial_yf\partial_xg\,$$
where $f=f_{m,s}\,x^{m+s+1}y^{s+1}$, $g=g_{m,t}\, x^{m+t+1}y^{t+1}$ ( implied
sum )
where $\{ \ast,\ast \} $ is the Poisson bracket of the coordinates $x$ and $y$
and $s,t\ge 0$.
By restricting the values of $s$ and $t$ to $0$ one recovers the
$w_2$ subalgebra of $w_\infty$.
This subalgebra admits a central extension and the algebra may be written as
$$ \big[ w_m,w_n \big]=(m-n)w_{m+n}+{c\over12}\,(m^3-m)\delta_{m+n}\ .$$
In [5] we showed that one may extract an effective action from almost any
infinite
dimensional Lie algebra by knowing the algebra's  two cocycle. That action is
$$
S=\int d\lambda\,d\tau\biggl(\langle k(g)\tilde f[\xi_\tau,\xi_\lambda]\rangle
+\tilde t c({\xi_\tau}_g,{\xi_\lambda}_g)\biggr), \eqno(1)$$
where the vector $\xi_\tau$ is used to denote a Hamiltonian vector field
generating changes along the time direction.  $\xi_\lambda$ is the Hamiltonian
vector field generating changes in a direction $\lambda$ which  corresponds to
a  one parameter family of group transformations from, say, the identity to
some group element $g(x,y)$~~ [11-13]. $k(g)\tilde f[\ast]$ is just
the coadjoint action of the group $G$ on the coadjoint vector $f$.  The
pure coadjoint orbit in the $w_2$ case  corresponds to an $SL(2,R)$
Chern-Simons theory, since $SL(2,R)$ is the isotropy group
of this coadjoint vector.

Let us take a moment to examine the $w_\infty$ algebra more closely.
As we just mentioned the $w_\infty$ algebra is usually written as the
Poisson bracket algebra of functions on a two-dimensional phase
space.
The $w_2$ central extension is easily added to this algebra to yield the
full centrally extended $w_\infty$ algebra.  Indeed for centrally extended
generators $\{w^i_m, \alpha_i\}$,
we write
$$\eqalign{ & \big[ \{w^i_m,\alpha_i\}  ,\{w^j_n,\alpha_j\} \big] \cr
& =\{ ((j+1)m-(i+1)n) w^{i+j}_{m+n} , {1\over 12}c(m^3-m) \delta_{m+n,0}
\delta^{i,0} \delta^{j,0} \}. \cr } $$
 The $\alpha_i$'s denote the generators for the central extensions of the
algebra and here only $\alpha_2$ is non zero.
{}From here we attempt to define the coadjoint representation of $w_\infty$.
We seek the generators $(\tilde w^l_m, \tilde \alpha_l)$ such that
$$ < \{\tilde w^l_k, \tilde \alpha_l\} \mid \{w^j_n,\alpha_j\} > = \delta^{l,j}
\delta_{n,k} + \tilde \alpha_l (\alpha_j) ,$$ where $\tilde \alpha_l (\alpha_j)
=
\delta^{l,2} \delta^{j,2}.$
 The action of the Lie algebra element $\{w^{(i)}_m,\alpha_i\}$
on  $\{\tilde w^l_k,\tilde \alpha_l\}$ is given by
$$\eqalign{ \{w^i_m,\alpha_i\} \big[ \{\tilde w^l_k &,\tilde \alpha_l \} \big]
\cr
&= -\{ ((l+2)m-(i+1)k) \tilde w^{l-i}_{k-m}
+{1\over12}c(m^3-m)\delta_{m,k-n}\delta^{0,l} \delta^{0,i} , 0 \}.
 \cr} \eqno(2)$$
Observe that if we demand that $i,l \ge 0$ the dual space or coadjoint
representation for this subalgebra is not  invariant under the action of the
adjoint representation.  Therefore one cannot
enforce this condition and also define a smooth dual for this algebra.  One
must allow {\it all} conformal spin field generators in order to describe the
symplectic geometry associated with the $w_\infty$ algebra.  Hence by using the
natural bilinear map which takes the unrestricted algebra to complex numbers we
can define the coadjoint representation.  In this sense the $w_\infty$ algebra
is more analogous to the
WZW type theories than two dimensional Polyakov gravity.  In other words,
recall that
the Kac-Moody algebras also enjoy the presence of a bilinear
symmetric form, where as Diff $S^1$ does not.  As we will see
shortly this problem persist for the $W_\infty$ algebra.
  One representation of the
symmetric form on $w_\infty$ (from here on we assume all conformal spins
are allowed) is given by  (~promoting x and y to complex variables~)
$$< w^i_m\mid w^j_n> = \oint \oint w^i_m w^j_n dx~dy. $$  The contour is a unit
circle about the origin.  From here it is easy to see that
$$ \tilde w^i_m= w^{1-i}_{-m}.$$

The action for $w_\infty$ was essentially constructed in Ref.[5] with the
exception that now
coadjoint vectors can come from full dual  of the area preserving
diffeomorphism algebra.   Observe that the pure covector, ${\bf B}=(0,\tilde
t)$, is invariant
with respect to all the generators except the $w_2$ generators (~modulo  {\bf
SL(2,R)}~)
which belong to ${\bf Diff}^{+}_0~{\bf R}^2$.
This orbit could correspond to the self-dual Einstein vacuum.
We will now focus on the $W_\infty$ algebra.

Recall that the $W_\infty$ algebra is defined by the commutation relations,
$$\big[ W^i_m,W^j_n \big] = \sum^{s=[j+k]/2}_{p=0} q^{2p} g^{jk}_{2p}(m,n)
W^{j+k-2p}_{m+n} + q^{2j} c_j m^{2j+3} \delta{j,k} \delta_{m+n,0}.$$
where $s$ is the maximal integer and the structure function have the form,
$$ g^{jk}_{2p}(m,n)=\phi^{jk}_{2p} N^{jk}_{2p}(m,n) $$
with
$$
\phi^{jk}_{2p}=\sum^p_{r=0} \prod^r_{l=1}{ (2l-3)(2l+1)(2p-2l+3)(p-l+1)\over
l(2j-2l+3)(2k-2l+3)(2j+2k-4r+2l+3)},$$and
$$
c_j={2^{2j}j!(j+2)!\over (2j+1)!! (2j+3)!!}c . $$

The $N^{jk}_{2p}(m,n)$ can be computed from the prescription described
below.
The structure functions,
$g^{ij}_{2r}$, vanish when $i+j-2r<0$ [9].  This guarantees that the algebra
will
terminate to
 generators with conformal spin two and higher.
However this is to no advantage in the construction of the coadjoint
representation.  Again we must allow the generators with conformal spin less
than two,
since only the $W_2$ subalgebras have an invariant dual space.
Let us construct the dual vectors, $ (\tilde W^l_m, \tilde \alpha_l) $, to this
algebra
such that $$  < \{\tilde W^l_k, \tilde \alpha_l\} \mid \{W^j_n,\alpha_j\} > =
\delta^{l,j} \delta_{n,k} + \tilde \alpha_l (\alpha_j).$$ Again the
$\alpha_l$'s and $\tilde \alpha_l$'s are the corresponding generators for the
$l$'th central extension in the adjoint and coadjoint vector spaces.  From the
invariance of this expression one finds that $$\eqalign{ \{W^i_m,\alpha_i\}
\big[ \{\tilde W^l_k &,\tilde \alpha_l \} \big] \cr
&=-\sum^{s={l\over2}}_{r=0} \{ g(m,k-m)^{i,l-i+2r}_{2r} \tilde W^{l-i+2r}_{k-m}
+  c_l(k) \delta^{i,l} \delta^{m,-k} , 0 \}.
 \cr} \eqno(3)$$ and we see that in order to define an invariant
dual space, we need to include all the conformal spin generators.  This
will technically facilitate matters, as we will see shortly.

In Ref.[4], Bakas has emphasized the importance of using a ``q'' bracket or
Moyal bracket [14] in the construction of the symplectic geometry for
generators
with conformal spin 2 and higher.  The Moyal bracket provides a suitable
deformation of the Poisson bracket that will permit Dirac quantization.
Ambiguities due to operator ordering are avoided.  In the limit as $\hbar
\rightarrow 0$, this bracket reduces to the Poisson bracket.
The  ``Lone Star'' $W_\infty$ [9] has (~at least formally~) a realization in
terms of the Moyal
bracket for all conformal spin fields [15].  Since we cannot restrict to
positive conformal spins,
 we will use the notation of Fairlie and Nuyts, Ref. [15], to describe this
algebra.  We define the Moyal bracket [14,15] between functions $f(x,y)$ and
$g(x,y)$ by,
$$
\{f,g\}_{{}_\hbar} = \sum^\infty_{p=0} (-1)^p{\hbar^{2p+1}\over
(2p+1)!}~\sum^{2p+1}_{k=0}(-1)^k \big( {2p+1\over k}\big)~(\partial^k_x
\partial^{2p+1-k}_y~f)(\partial^{2p+1-k}_x \partial^k_y g) \eqno(4)$$
Then, using this bracket, one may obtain a renormalized $W_\infty$ algebra by
using the
generators $$W^j_m=i \lambda \exp({i m x\over y}) y^{2(j+1)}, \eqno(5)$$ with
$i,j\ge 0$.  This algebra
admits central extensions for all positive conformal spins.  We write
the centrally extended $W_\infty$ algebra as ( here the $i=0$ term
will correspond to the spin two components )
$$
\big[ W^i_m,W^j_n \big] = \sum_{p=0} q^{2p} N^{jk}_{2p}(m,n) W^{j+k-2p}_{m+n} +
q^{2j} c_j m^{2j+3} \delta{j,k} \delta_{m+n,0}.$$
Note that from the action of $W^0_m$ on $W^i_n$ that the generators do
not have definite conformal spin.
The structure functions, $N(m,n)^{jk}_{2p}$, can be computed directly using
the Moyal bracket and are the same as those used by Pope, Romans, and Shen in
Ref.[9].  By enforcing the Jacobi identity one finds that
$$ c_{j+1}=c {(2)^{j+1} \over (j+1)! }.$$   The duality condition for the
representation given in Eq.(5) is
$$\eqalign{
 & < \{\tilde W^l_k, \tilde \alpha_l\} \mid \{W^j_n,\alpha_j\} >= \cr
&{1\over 4\pi^2} \oint dy \int^{2\pi}_0 \big(-i \lambda \exp({- i l x\over y})
y^{-2k-3}\big)\big( i \lambda \exp({i n x\over y}) y^{2(j+1)}\big) dx +
 \tilde \alpha_l(\alpha_j),\cr}$$
where care is taken to do the $x$ integration first.
  Following Eq.(1) we may write down the action for $W_\infty$.
For the  generators given in Eq.(5), the  two cocycle for the renormalized
$W_\infty$ is given by
$$
c(f,g)=\sum^\infty_{i=0} {c_i (-1)^{i+1}\over 4\pi^2 (2(i+1)!^2}  \int \oint
D_y^{2(i+1)} \partial^{i+1}_x( y^{i+1}f)~{1\over
y}~D_y^{2(i+1)}\partial^{i+2}_x(y^{i+2}g) ~dx~dy,
$$ where the derivative operator $D_y~=\partial_y + {x\over y}\partial_x .$
Instead of using the Hamiltonian vector fields we may use the Hamiltonians
corresponding to generators in the direction of time, $\tau$, and the group
direction, $\lambda$.  Using the Moyal bracket and its correspondence principle
[16] we can
identify the canonical
variables as $x$ and $y$.  Then given an area preserving diffeomorphism
$$ x\rightarrow \varphi_1(x,y), ~~~~~~~y\rightarrow \varphi_2(x,y), $$
For the generators of time translations and group transformations we
 may write $$\partial_b H_\tau = {\partial_\tau \varphi^\alpha (x,y)
\over \partial_a \varphi^\alpha}
 \epsilon_{ab}, ~~\partial_b H_\lambda= {\partial_\lambda \varphi^\alpha (x,y)
\over \partial_a \varphi^\alpha}
 \epsilon_{ab}$$
respectively.
Here $a,b \in \{1,2\}$.
 The $W_\infty$
effective action for the orbit corresponding to the coadjoint vector
$ {\bf B}=(\tilde B, t^l \tilde \alpha_l )$ is
$$\eqalign{
&S(\tilde B)= \cr
& \sum^\infty_{i=0} t^i {c_i\over 4\pi^2}  \oint\int^{2\pi}_{0}\int\int
D_y^{2(i+1)}\partial^{i+1}_x\big(y^{i+1}H_\tau\big)~D_y^{2i+2}
\partial^{i+2}_x\big(y^{i+2}H_\lambda\big) ~dx~dy~d\lambda~d\tau \cr
& +  \oint\int^{2\pi}_0\int\int~ \tilde B(x,y)\{H_\tau,
H_\lambda\}_{{}_{\hbar}}~dx~dy~d\lambda~d\tau. \cr} \eqno(6)
$$

 First notice that  the coordinates
in Eq.(4) represent the algebra on $S^1\otimes R$.  As long as one is
interested in only the algebraic structure this compactification offers
no significant changes as compared to $R^1~\otimes~R^1$. However in our case we
are actually interested in group transformation on coadjoint orbits.
One may recall [12,13] that the isotropy groups for the orbits in Polyakov
gravity
depend on whether we specify two dimensional gravity on $R^2$
or $S^1\otimes R$.  In the first case, the pure covector corresponds to the
field space (~orbit~)$ {{\bf Diff}S^1/ Sl(2,R)}$
and in the second case to ${ {\bf Diff}S^1/ S^1}$.  The action
is a $2+1$ space-time action with one more dimension for the group direction.
We have left it in the four dimensional form since WZ terms are present.
The existence of such terms will require that $t^i c_i$ be integers.
Since this action arises from four dimensional theories
with a Killing symmetry, the number of dimensions is correct.  Recently
a two dimensional realization of $W_\infty$ has been found  for a free
complex bosonic field [17].

One may characterize the orbits by the subalgebras that
leave {\bf B} invariant.   By knowing the available subalgebras one can work
backwards to find representative covectors for each orbit.
In this case the pure covector,
${\bf B}=(\tilde B = 0, \sum_{i=0} t^ic_i )$, will certainly be invariant
under ${\bf Diff}^{-}_0~~ {\bf R}^2$ and the Abelian subgroup generated
by $W^{-1}_k$.  Thus ${{\bf Diff}~~ {\bf R}^2\over {\bf Diff}^{-}_0~~ {\bf R}^2
\otimes H }$ is the field space and also corresponds to vacuum solutions.
Non-pure
orbits could be realized as cosmological terms coupled to gravity and
higher spin tensor fields.

Some interesting question emerge from this action.  In the case of $w_\infty$,
Park has shown that the equations of motion of the action
$$S(u) = \int ({1\over2} u {\bar \partial} \partial u +  \partial_z^2
\exp{u})dx~dy~dz,$$
corresponding to self-dual Einstein solutions with one Killing symmetry, have
$w_\infty$ symmetry.  Indeed we have tried to
argue that coupling
the Eguchi-Hanson instanton to spectating Rarita-Schwinger
fields yields the $W_2$ effective action. This further suggests that the pure
central orbit of $W_2$ may correspond to a bosonization prescription of
the Rarita-Schwinger field on a self-dual manifold with one Killing
vector field.
By working backwards from the  $W_\infty$ algebra we may be able to
reconstruct instantons  with hidden $W_\infty$
symmetry.  We expect the  $W_\infty$ action
to be a consistent coupling of higher spin fields to a self-dual(?)
four dimensional, linearly deformed, gravitational instanton.
This is reminiscent of  string theories, since a consistent string theory  also
requires
an infinite tower of coupled fields.
This may be a realistic expectation of quantum gravity in four dimensions.

Also, recall that one may define the conformal spin of a field through the
action of the Virasoro subalgebra of these $W$ algebras [18].  Indeed
in a one dimensional realization of $w_2$ we know that,
$$ w^0_k (z^{m+1}) dz^{\Delta}=-((m+1)+\Delta (k+1))z^{m+k+1} dz^{\Delta}.$$
In the case of quantum two dimensional Polyakov gravity [18,10-13], the
adjoint vectors correspond to conformal spin -1 operators and
the covectors dual to them are the quadratic differentials.  The importance of
the
quadratic differentials in the quantization of two dimensional gravity is that
they are the pseudo-metrics corresponding to conformally gauge fixed metrics,
	$ ds^2=g_{ab}dx^a dx^b$.  Here there is a  clear distinction between the
contravariant and covariant vectors in terms of the conformal spin.
Because we required all conformal spin generators for the construction
of the symplectic structure of the
 $w_\infty$  and $W_\infty$ algebras , the distinction between derivative
operators
and the fields can no longer be determined just be checking the
conformal spin.  In other words  the higher order
differentials can be related to ``conformally'' gauge fixed fields, $i.e.$ $~~
ds^3= A_{abc}dx^a dx^b dx^c$, and be assigned positive conformal spin.
However, in  the $w_\infty$ and $W_\infty$ algebras, the familiar ``in'' and
``out'' states as mentioned in the original work of Belavin, Polyakov, and
Zamolodchikov [18] are dual to each other, but there are two spaces of {\it
all} the conformal spins that are respectively dual.

Let us remark on one more observation.  We have seen that there are two
separate algebras that contain either one or an infinite number
of central extensions.  It has been argued that both are
related to the area preserving diffeomorphism in 2D.  As far as we
know $w_\infty$ and $W_\infty$ are not unitarily related.  Furthermore
it is well known that there may be many different ways to take the
limit of the Zamolodchikov $W_N$ algebras as $N \rightarrow \infty$
and recover a Lie algebra.

Then how do these ambiguities translate into the four dimensional theories?
We suspect  that the four dimensional self-dual manifolds support
instantons, which arise from different limits, are not diffeomorphic to each
other.  Recall that for compact
four manifolds, the diffeomorphism class is specified by the cup
product of $H^2(M^4,Z)$ with itself [20].  In other words, the intersection of
the two
manifolds associated with the homology classes determines what functions,
if any, will be smooth on the four manifold.    Even in the case
of $R^4$, there can be many (~at least two-fold uncountably infinite~)
inequivalent notions of a  ring of smooth functions.  Roughly speaking, one
surgers a hole
about the topology of a compact manifold that does not admit a ring of
smooth functions to produce an open four manifold which is homeomorphic
 but not diffeomorphic to $R^4$, yet still admits a smooth ring of functions.
To add further relationships between 2D and 4D, recall that
on compactified Euclidean space-times that all ( stable, finite ) solutions
of the pure Yang-Mills equations are self-dual [21].  This together with the
results of [2] and [3] tends to support the fact that 2D conformal theories may
indeed drive the differential structure of theories in four-dimensions.
We are presently investigating these issues.

{ Acknowledgements}:  We thank Didier Depireux, Ralph Lano and
Yannick Meurice for discussion.

\vfill\eject
\nopagenumbers
\baselineskip=13pt
\centerline{\bf REFERENCES}
\settabs 1 \columns
\+ [1]  A.A. Belavin, A.M. Polyakov and A.B. Zamolodchikov, Nucl. Phys B 241
(1984) 333 \cr
\+ [2]  M. Atiyah, Comm. Math. Phys. 93 (1984) \cr
\+ [3]  Q.-H. Park, Phys. Lett. B238 (1990) 287; \cr
\+        Phys.Lett. B236 (1990) 429; \cr
\+ {\it Self-Dual Yang-Mills (+ Gravity) As A  2-D SigmaModel}, UMD Preprint
\cr
\+ [4]  I. Bakas, Phys. Lett. B228 (1989) 57; \cr
\+ Comm. Math. Phys. 134, (1990) 487; \cr
\+ [5]  V.G.J. Rodgers, {\it A $W_2$ Effective Action }, to appear in
Mod. Phys. Lett. A \cr
\+ [6]  T. Eguchi and A.J. Hanson, Phys. Lett. B74 (1978) 249 \cr
\+      C. Boyer and J. Finley, J. Math. Phys. 23 (1982) 1126 \cr
\+      J. Gegenberg and A. Das, Gen. Rel. Grav. 16 (1984) 817 \cr
\+ [7]  A.J. Hanson and H. R\"omer, Phys. Lett. B80 (1978) 58 \cr
\+ [8]  S.W. Hawking and C.N. Pope, Nucl.Phys.B146 (1978) 381 \cr
\+ [9]  C.N. Pope, L.J. Romans and X.Shen,  Phys. Lett B236 (1990) 173 \cr
\+ [10]  A.Bilal, Phys. Lett. B227 (1989) 406 \cr
\+ [11] F. Zaccoria, E.C.G. Sudarshan, J.S. Nilsson, N. Mukunda, G. Marmo,\cr
\+ and A.P. Balachandran,  Phys. Rev D27 (1983) 2327;\cr
\+ A.P. Balachandran, G. Marmo, B.S. Skagerstam, and A. Stern,\cr
\+ {\it Gauge Symmetries and Fibre Bundles; Applications to Particle
Dynamics},\cr
\+ Springer-Verlag Berlin (1983)\cr
\+ [12]  B. Rai and V.G.J. Rodgers, Nucl Phys. B341 (1990) 119 \cr
\+       A. Yu Alekseev and S.L. Shatashvili, Nucl. Phys. B323 (1989) 719\cr
\+ [13]  G.W. Delius, P. van Nieuwenhuizen, and V.G.J. Rodgers,\cr
\+ Inter. Jour. of Mod. Physics A5 (1990), 3943 \cr
\+ [14]  J. Moyal, Proc. Camb. Phil. Soc. 45 (1949)  99 \cr
\+ [15]  D.B. Fairlie and J. Nuyts,  Comm. Math. Phys. 134, 413-419 (1990)\cr
\+ [16]  T.F. Jordan and E.C.G. Sudarshan, Rev. Mod. Phys. 33 (1961) 515 \cr
\+ [17]  I. Bakas and E. Kiritsis Nucl.Phys.B343 (1990) 185 \cr
\+ [18]  A.B. Zamolodchikov, Teor. Math. Fiz. 65 (1985) 347; \cr
\+ Theor. Math. Phys. 65(1986) 1205 \cr
\+ [19]  A.M. Polyakov, Mod. Phys. Lett. A2 (1987) 893 \cr
\+ [20]  D. Freed and K. Uhlenbeck, ``Instantons and Four Manifolds'',\cr
\+ Second Edition, ( Spinger-Verlag, 1991 ) \cr
\+ [21]  J.P. Bourguignon, H.B. Lawson, and J. Simon, Proc. Nat. Acad. Sci 76
(1979) 1550 \cr

\end